\documentclass{aa}
\begin{document}
\title{Generation of the magnetic field in jets}
\author{V.~Urpin}
\offprints{V.~Urpin}
\institute{Departament de F\'{\i}sica Aplicada, Universitat d'Alacant,
           Ap. Correus 99, 03080 Alacant, Spain \\ 
           A.F.Ioffe Institute of Physics and Technology and Isaak 
           Newton Institute of Chili, Branch in St.Petersburg, 
           194021 St.Petersburg, Russia}

\date{today}

\abstract{We consider dynamo action under the combined influence of 
turbulence and large-scale shear in sheared jets. Shear can stretch 
turbulent magnetic field lines in such a way that even turbulent 
motions showing mirror symmetry become suitable for generation 
of a large-scale magnetic field. We derive the integral induction
equation governing the behaviour of the mean field in jets. The 
main result is that sheared jets may generate a large scale
magnetic field if shear is sufficiently strong. The generated mean 
field is mainly concentrated in a magnetic sheath surrounding the 
central region of a jet, and it exhibits sign reversals in the 
direction of the jet axis. Typically, the magnetic field in a sheath 
is dominated by the component along the jet that can reach 
equipartition with the kinetic energy of particles. The field 
in the central region of jets has a more disordered structure.  
}

\titlerunning{Generation of the magnetic field in jets}
\maketitle

\section{Introduction}

It is widely believed that magnetic fields play an important
role in the formation and propagation of astrophysical jets providing 
an efficient mechanism of collimation through magnetic tension forces 
(see, e.g., Hughes 1991, Blandford 1993, K\"{o}nigl \& Pudritz 1999). 
Polarization observations provide information on the orientation and 
degree of order of the magnetic field in jets. It appears that many jets 
can develop relatively highly organized magnetic structures 
(see, e.g., Cawthorne et al. 1993, Lepp\"{a}nen, Zensus \& Diamond 
1995, Gabuzda 1999). For example, radio emission from the jets of 
NGC 4258 indicates that the magnetic field is oriented mainly along 
the jet axis but a noticeable toroidal component can also be presented 
(Krause \& L\"{o}hr 2004). The conventional estimate of the magnetic 
field strength  in jets comes from the minimum energy argument and 
corresponds to approximate equipartition between magnetic and jet 
particle energy (see, e.g., Laing 1993). Recent observations of jets
in TeV BL Lac objects confirm this estimate (Ghisellini, Tavecchio,
\& Chiaberge 2005). However, Kataoka \& Stawarz (2005) 
argue that the powerful jets in quasars and FR II objects can be far 
from the minimum energy condition, and the field strength very likely 
exceeds the equipartition value. Upper limits to the inverse Compton 
radiation of the jet in M87 imposed by HESS and HEGRA Cerenkov 
Telescopes also indicate that the magnetic field cannot be weaker than 
the equipartition value (Stawarz et al. 2005).

To explain the observational data, various simplified models of  
three-dimensional magnetic structures have been proposed (see, e.g., 
Chan \& Henriksen 1980, Laing 1981, 1999). For example, Canvin et al. 
(2005) computed emission and polarization for a few different models of 
the jets in NGC 315 and concluded that both fully and partially 
ordered fields can produce the high fractional polarization observed. 
It seems (Laing 1999) that good agreement with observations can be 
obtained if the jet is considered as a cylindrical core with more or 
less uniform density and pressure surrounded by a shear layer. Possibly, 
the core and shear layer are both turbulent (see, e.g., Gabuzda 1999). 
The magnetic field has a substantial longitudinal component in the 
shear layer, but the field can be random or a transverse component 
dominates in the core region. Recent VLBI observations (Hirabayashi 
et al. 1998, Gabuzda, Murray, \& Cronin 2004) indicate that the 
transverse field may have a substantial toroidal component in some 
objects. 

The mechanisms responsible for generation of the magnetic field in 
jets are still unclear. Since the origin of jets is probably relevant 
to MHD-processes in magnetized plasma, their magnetic fields could be 
generated during the process of jet formation (see, e.g., Blandford 
\& Payne 1982, Romanova \& Lovelace 1992, Koide, Shibata \& Kudoh 1998). 
However, as mentioned, jets are possibly turbulent, and turbulent 
dissipation should lead to a rapid decay of any initially ordered fields 
if the mean-field dynamo does not operate. Besides, the toroidal 
component becomes dominant as the flow expands (Begelman, Blandford \& 
Rees 1984) and, under certain conditions, this component can be unstable 
thereby decollimating the jet (see, e.g., Spruit, Foglizzo \& Stelhe 
1997, Begelman 1998). Therefore, some mechanisms for generating the 
magnetic field should certainly be operative in jets. One of these 
possibilities has been considered by Honda \& Honda (2002), who argued 
that the stream can generate a toroidal field that participates in 
self-pinching the plasma in a fully relativistic jet consisting of 
electrons, positrons, and a small portion of ions. 

The magnetic field in jets can also be generated by some dynamo
mechanism. It is plausible that jets are turbulent making the 
turbulent dynamo a suitable candidate. This point is argued by
Stawarz et al. (2005) for the particular case of the jet in M87 where
the field strength is probably stronger than the equipartition value. 
The origin of turbulence can be attributed to different instabilities 
arising in jets. The classical Kelvin-Helmholtz instability seems to 
be the dominating factor of destabilization in the most simplified jet
model with one bulk velocity and a narrow interface with the external 
medium. This instability has been studied by many authors both in 
linear (e.g., Blandford \& Pringle 1976, Ray 1981, Payne \& Cohn 1985, 
Zhao et al. 1992) and  nonlinear regimes (Bodo et al. 1994, Koide, 
Nishikawa \& Mutel 1996, Nishikava et al. 1997, Hardee et al. 1998). 
In more refined models, however, other hydrodynamic instabilities can 
manifest themselves leading to production of turbulence in jets 
(Urpin 2002a, Alloy et al. 2002). Gvaramadze et al. (1984) proposed 
the model where the field is generated due to the combined influence 
of helical turbulence and regular flow that stretches a seed magnetic 
field. In this model, however, the field grows superexponentially only
during the initial stage but decays eventually, sothat a true dynamo is 
replaced by a temporal growth of the field. 

In this paper, we consider the turbulent dynamo action that can be 
responsible for generation of the magnetic field in jets. Our model
is based on the shear-driven dynamo action associated with turbulent 
shear flows (see, e.g., Urpin 1999, 2002b). Shear stretches turbulent 
magnetic field lines in the direction of a mean flow, which in turn 
generates the additional component of the mean electromotive 
force, which is proportional to the production of shear and magnetic 
field. Plausibly, the flow inside jets is sheared and turbulent, and 
the shear-driven dynamo can be in action there. There is observational 
evidence that the jet structure is rather complex and that different bulk 
velocities can be represented inside the jet; for more details see 
the discussion in Hanasz \& Sol (1996) and Stawarz \& Ostrowski (2002). 
Theoretical models of the jet formation also predict the existence of 
a nonvanishing transverse gradient in the jet velocity (see, e.g., 
Melia \& K\"{o}nigl 1989, K\"{o}nigl \& Kartje 1994, Sol, Pelletier 
\& Asseo 1989).

In the present paper, we show that the shear-driven dynamo can 
naturally explain the observed magnetic structure of jets. In Sect.2, 
we derive the main equations governing the magnetic field in sheared 
turbulent jets. The geometry and growth rate of the generated field
is then considered in Sect. 3. A brief summary of our results is 
finally presented in Section 4.  

\section{The mean-field electrodynamics of jets}

We model the jet as a cylindrical flow in the $z$-direction with 
radius $R$. Plasma inside the jet has a velocity $\vec{V}= V(r) 
\vec{e}_{z}$, and $r$, $\varphi$, $z$ are the cylindrical coordinates with 
$\vec{e}_{r}$, $\vec{e}_{\varphi}$, and $\vec{e}_{z}$ the 
corresponding unit vectors. For the sake of simplicity, we assume that 
the jet velocity is subrelativistic. Without loss of generality, we 
can assume that density $\rho$ is constant inside the jet. The 
Reynolds number is large in jets and, possibly, the plasma of sheared jets 
is turbulent. Variations in the flux, which are observed in many 
jets, are sometimes interpreted in terms of this turbulence (e.g., 
Marscher, Gear \& Travis 1992, Massaro et al. 1999).

We can represent the magnetic field $\vec{\cal B}$ and the velocity 
$\vec{u}$ as a sum of the mean and fluctuating parts, $\vec{\cal B} = 
\vec{B} + \vec{b}$ and $\vec{u}= \vec{V} + \vec{v}$, where $\vec{B}$ 
and $\vec{V}$ are the mean field and velocity, respectively. We neglect 
dissipative effects in the induction equation and assume the field to 
be frozen in plasma. Then, averaging the induction equation, we have 
for the mean field
\begin{equation}
\frac{\partial \vec{B}}{\partial t} = \nabla \times (\vec{V} 
\times \vec{B}) + \nabla \times 
\vec{\cal E},
\end{equation}
where 
\begin{equation}
\vec{\cal E} = \langle \vec{v} \times \vec{b} \rangle
\end{equation}
is the mean electromotive force; $\langle...\rangle$ denote ensemble 
averaging. In what follows, however, we will need the sign of 
dissipative terms to choose the integration path properly when
calculating Fourier integrals. Generally, a weak dissipation in jets 
can be provided by either electrical resistivity (if plasma is 
collisional) or some plasma mechanisms, such as Landau 
damping (if plasma is collisionless). 

We consider the mean electromotive force $\vec{\cal E}$ using a 
quasilinear approximation. In this approximation, mean quantities are 
governed by equations including non-linear effects in fluctuating 
terms, whilst the linearized equation is used for the fluctuating 
quantities (Krause \& R\"{a}dler 1980). A quasilinear approximation is 
accurate enough, for example, to describe an ensemble of waves 
with relatively high frequencies and small amplitudes. Then, the 
linearized induction equation for the fluctuating magnetic field 
reads
\begin{equation}
\frac{\partial \vec{b}}{\partial t} = \nabla \times ( \vec{V} 
\times \vec{b}) + \nabla \times (\vec{v} \times \vec{B} ).
\end{equation}
If $\vec{V}= V(r) \vec{e}_{z}$, then we have
\begin{equation}
\frac{\partial \vec{b}}{\partial t} + V(r) \frac{\partial 
\vec{b}}{\partial z} - \vec{e}_{z} b_{r} V'(r) = \vec{A} , 
\end{equation}
where
\begin{equation}
\vec{A} = (\vec{B} \cdot \nabla) \vec{v} - (\vec{v} \cdot 
\nabla) \vec{B} - \vec{B} \nabla \cdot \vec{v},
\end{equation}
and $V' = d V/ d r$. 

Equation(4) can be solved by making use of a partial Fourier 
transformation. Since coefficients in Eq.~(4) do not depend on 
$\varphi$ and $z$ we make initial transformations in these 
coordinates. The fluctuating magnetic field can be represented as
$$
\vec{b}(\vec{r}, t) = \sum_{m} \int d k_{z} \hat{\vec{b}}_{1}(r,
m, k_{z}, t) e^{- i m \varphi - i k_{z}z}, 
$$
where $m$ is an integer. Then,  
\begin{equation}
\hat{\vec{b}}_{1}(r, m, k_{z}, t) = \frac{1}{(2 \pi)^{2}}
\int d \varphi d z e^{i m \varphi + i k_{z} z} 
\vec{b}(\vec{r}, t),
\end{equation}
and the equation for $\hat{\vec{b}}_{1}$ reads
\begin{equation}
\frac{\partial \hat{\vec{b}}_{1}}{\partial t} - i k_{z} V 
\hat{\vec{b}}_{1} - \vec{e}_{z} V' \hat{b}_{1r} =
\hat{\vec{A}}_{1},
\end{equation}
where $\hat{\vec{A}}_{1}(r, m, k_{z}, t)$ is the corresponding 
Fourier amplitude of $\vec{A}(\vec{r}, t)$. Substituting 
$\hat{\vec{b}}_{1} = e^{i k_{z} V t} \hat{\vec{b}}_{2}$, we obtain 
the equation for $\hat{\vec{b}}_{2}$ that does not contain the 
advective term,
\begin{equation}
\frac{\partial \hat{\vec{b}}_{2}}{\partial t} - \vec{e}_{z} V'
\hat{b}_{2r} = e^{-i k_{z} V t} \hat{\vec{A}}_{1}.
\end{equation}
This equation can be solved by Fourier transformation in $t$.
Introducing
\begin{eqnarray}
\hat{\vec{b}}(r, m, k_{z}, \omega) = \frac{1}{2 \pi} \int dt
e^{-i \omega t} \hat{\vec{b}}_{2}(r, m, k_{z}, t)
\nonumber \\
= \frac{1}{2 \pi} \int dt e^{- i(\omega + k_{z} V) t} \hat{\vec{b}}_{1}
(r, m, k_{z}, t), 
\end{eqnarray}
we obtain the following expression from Eq.~(9) 
\begin{equation}
\hat{\vec{b}} = - \frac{i}{\omega} 
\hat{\vec{A}}(r, m, k_{z}, \omega) - \frac{V'}{\omega^{2}} 
\; \hat{A}_{r}(r, m, k_{z}, \omega) \vec{e}_{z}.
\end{equation}
Expression (10) is not a complete Fourier transform of 
$\vec{b}(\vec{r}, t)$ since it depends on the radial coordinate. Note 
that, in reality, Eq.~(10) does not contain singularities because 
neglected dissipative terms would result in small negative imaginary 
corrections to $\omega$, so we would have $\omega - i0$ instead of 
$\omega$ in singular terms. Here, $\pm i0$ denotes a positive (or 
negative) 
contribution to $\omega$ caused by weak dissipative effects. The sign 
of this small imaginary term is important when calculating Fourier 
integrals. 

The solution for a fluctuating magnetic field reads  
\begin{eqnarray}
\vec{b}(\vec{r}, t) = \sum_{m} \int \frac{d 
\omega d k_{z}}{i \omega} e^{i (\omega + k_{z} V)t - 
i m \varphi - i k_{z} z} 
\nonumber \\
\times \left( \hat{\vec{A}} - 
\frac{i V'}{\omega} \; \hat{A}_{r} \vec{e}_{z} \right).
\end{eqnarray}
Substituting this expression into the definition of $\vec{\cal E}$,
we obtain
\begin{eqnarray}
\vec{\cal E} = \sum_{m} \int \frac{d \omega d k_{z}}{i (2 \pi)^{3}
\omega} d \varphi_{1} dz_{1} dt' e^{i\omega (t-t')   - i k_{z}[z-z_{1}
-V(t-t')]}
\nonumber \\
e^{-im(\varphi - \varphi_{1})} \langle \vec{v}(\vec{r}, t)
\times \left[ \vec{A}(\vec{r}_{1}, t') - \frac{i V'}{\omega} 
A_{r}(\vec{r}_{1}, t') \vec{e}_{z} \right] \rangle,
\end{eqnarray}
where $\vec{r}_{1} = (r, \varphi_{1}, z_{1})$. Taking into account that 
summation over $m$ and integration over $d k_{z}$ yield the corresponding
$\delta$-functions, we can simplify expression (12):
\begin{eqnarray}
\vec{\cal E} = \int \frac{d \omega dt'}{2 \pi i \omega} e^{i \omega 
(t-t')} \langle \vec{v}(\vec{r}, t) \times \left[ \vec{A}(\vec{r}', t') 
\right.
\nonumber \\
\left.
- \frac{i V'}{\omega} A_{r}(\vec{r}', t') \vec{e}_{z} \right] \rangle
\mid_{\vec{r}' = \vec{r} - \vec{V}(t-t')}. 
\end{eqnarray}
Integrals over $d \omega$ can now be calculated using the known 
integrals if we note that $\omega$ has a small imaginary part caused by 
dissipation. We have (see Gradshtein \& Ryzhik 1965 
\begin{equation}
\frac{1}{2 \pi} \int_{-\infty}^{\infty} \frac{e^{- i px}dx}{(ix +
\beta)^{\nu}} = \frac{(-p)^{\nu -1}}{\Gamma(\nu)} e^{\beta p} \;\;\;\; 
{\rm if}
\;\; p<0,
\end{equation}
and $0$ if $p>0$. The parameter $\beta$ in this integral is small in our
model because it is caused by dissipation. Then, the expression for 
$\vec{\cal E}$ transforms into
\begin{eqnarray}
\vec{\cal E} = \int_{-\infty}^{t} dt' \langle \vec{v}(\vec{r}, t) \times 
\left[ \vec{A}(\vec{r}', t') + \right. 
\nonumber \\
\left. (t-t') V' A_{r}(\vec{r}', t') \vec{e}_{z}
\right] \rangle \mid_{\vec{r}' = \vec{r} - \vec{V} (t-t')}.
\end{eqnarray}

Since $\vec{A}(\vec{r}', t')$ depends on the turbulent velocity 
(see Eq.~(5)) we can now calculate $\vec{\cal E}$ if specifying the 
correlation properties of turbulence. For the sake of simplicity, we 
assume turbulence to be locally isotropic and homogeneous with the 
correlation tensor given by
\begin{equation}
\langle \hat{v}_{i}(\omega', \vec{k}') \hat{v}_{j}(\omega'', \vec{k}'') 
\rangle = P  k_{i}' k_{j}'  
\delta(\vec{k}' + \vec{k}'') \delta(\omega' + \omega''),
\end{equation}
where $P=P(\omega', \vec{k}')$ is the spectral function and $(i, j)$ 
denote Cartesian components. This correlation tensor corresponds to 
acoustic turbulence (see, e.g., R\"{u}diger 1989), which seems to be  
plausible in a supersonic jet flow. For instance, sound waves can be 
generated by the Kelvin-Helmholtz instability at the jet boundary and 
then propagate through the jet volume generating fluctuations of 
the velocity and density (Payne \& Cohn 1985). 

Since correlation tensor (16) is of particular simplicity in
Cartesian components, it is convenient to represent the turbulent 
velocities $\vec{v}(\vec{r}, t)$ and $\vec{v}(\vec{r}', t')$ 
in Eq.~(15) in terms of Fourier integrals with Cartesian 
wavevectors as
$$
\vec{v}(\vec{r}, t) = \int d \omega' d \vec{k}' e^{i \omega' t -
i \vec{k}' \vec{r}} \hat{\vec{v}}(\vec{k}', \omega').
$$
Substituting this expression into Eq.~(15), we obtain, after 
ensemble averaging, 
\begin{eqnarray}
\vec{\cal{E}} =  \int_{-\infty}^{t} d t' \int d \omega' d \vec{k}'  
e^{i \omega' (t - t') -i \vec{k}' (\vec{r} -
\vec{r}')} P(\omega', \vec{k}') 
\nonumber \\
\times [ \vec{E}_{1} - (t-t') V' \vec{e}_{z} \times \vec{E}_{2} ] ,  
\end{eqnarray}
where
$$
\vec{E}_{1} = -\vec{k}' \times [ i k'^{2} \vec{B} +
(\vec{k}' \cdot \nabla') \vec{B} ] ,
$$
$$
\vec{E}_{2} =  \vec{e}_{z} \times \vec{k}'
\{ i k'_{r} (\vec{k}' \cdot \vec{B}) - i k'^{2} B_{r} -
[(\vec{k}' \cdot \nabla' ) \vec{B}]_{r} \},
$$ 
and $\nabla'= (\partial/\partial r', \partial/r \partial \varphi', 
\partial/ \partial z')$. Note that $\vec{B}$ in these expressions is 
a function of $\vec{r}'$ and $t'$.

Since turbulence is locally isotropic and homogeneous in a co-moving 
frame, the spectral function $P(\omega', \vec{k}')$ should be an even 
function of the frequency $\omega''= \omega' - k_{z}' V$ measured in 
a co-moving frame, i.e. we have 
$P(\omega'- k_{z}' V, \vec{k}')= P(-\omega' + k_{z}' V, \vec{k}')$. 
Then, denoting $P(\omega' - k_{z}' V, \vec{k}')= G(\omega'', \vec{k}')$, 
we can transform Eq.~(17) into
\begin{eqnarray}
\vec{\cal E}= \int d \omega'' d \vec{k}' G(\omega'',
\vec{k}') \int_{- \infty}^{t} dt' e^{i \omega'' (t-t')}
\nonumber \\
\times  [\vec{E}_{1}(\vec{r}', t') - (t-t') V' \vec{E}_{2}(\vec{r}', 
t')] |_{\vec{r}'=\vec{r}- \vec{V}(t-t')}.
\end{eqnarray}
Averaging of $\vec{E}_{1}$ and $\vec{E}_{2}$ over directions of 
$\vec{k}'$ yields
\begin{equation}
\vec{E}_{1} = - \frac{k'^{2}}{3} \; \nabla' \times 
\vec{B}, \;
\vec{E}_{2} = - \frac{k'^{2}}{3} \; \vec{e}_{z} \times (\nabla' 
\vec{B})_{r},
\end{equation} 
where $(\nabla' \vec{B})_{r}= \nabla' B_{r} - \vec{e}_{\varphi} 
B_{\varphi}/r'$. Finally, the expression for the mean electromotive 
force reads
\begin{eqnarray}
\vec{\cal{E}} = - \int_{-\infty}^{t} dt' F(t-t') [ \nabla' \times
\vec{B} 
\nonumber \\
- (t-t') V' \vec{e}_{z} \times (\nabla' \vec{B})_{r}]|_{\vec{r}'=
\vec{r}  - \vec{V} (t-t')},
\end{eqnarray}
where
\begin{equation}
F(t) = \frac{4 \pi}{3} \int d \omega dk k^{4} G(\omega, \vec{k})
e^{i\omega t}. 
\end{equation}
Note that the mean electromotive force given by Eq.~(20) is nonlocal 
in our approach since the turbulent magnetic field is determined by 
its previous evolution under the influence of shear.

\section{Generation of a large-scale field in jets}

We adopt the simplest model of a jet assuming that shear is 
relatively weak in the central region but is stronger in a shear 
layer near the jet surface, $r=R$. If $\Delta r$ is the thickness of 
a shear layer, then the shear-driven dynamo operates in the region 
$R \geq r \geq R- \Delta r$. Simulations indicate that often the 
thickness of a region with strong shear can be much smaller than the 
jet radius (see Alloy et al. 1999a,b). At that point, we will not 
specify the radial dependence of $V(r)$ because this dependence seems 
to be rather uncertain from both theoretical and observational 
points of view. However, we show that the proposed mechanism can 
generate the magnetic field for any dependence $V(r)$. 

In a stationary jet, the solution of the mean induction equation (1)
can be represented as
\begin{equation}
\vec{B}(\vec{r}, t) = \vec{B}(r) e^{\gamma t- i K_{z} z - i M \varphi},
\end{equation}
where $K_{z}$ is a wavevector of the magnetic field in the 
$z$-direction, $M$ the azimuthal wavenumber, and $\gamma$ the 
growth rate. Solution (22) describes spiral magnetic waves. 

We are particularly interested in the generation of large-scale fields 
with not very large $M$. Since the thickness of a shear layer is 
typically smaller than $R$, we can neglect terms of the order of $1/r$ 
compared to $\partial /\partial r$ in Eq.~(1). Then, the $r$-component 
of Eq.~(1) reads 
\begin{eqnarray}
\frac{\partial B_{r}}{\partial t} + V(r) \frac{\partial 
B_{r}}{\partial z} =
\int_{0}^{\infty} d \xi F(\xi) 
\left[ \Delta' B_{r}(\vec{r}', t-\xi)  \right.
\nonumber \\
\left. - \xi V'(r) 
\frac{\partial^{2}}{\partial r' \partial z'} 
B_{r}(\vec{r}', t-\xi) \right]_{\vec{r}'=\vec{r}- \vec{V} \xi},
\end{eqnarray}    
where $\Delta'$ is the Laplacian with the primed coordinates. 
Substituting dependence (22) for $B_{r}(r)$, we
obtain
\begin{eqnarray}
[ \gamma- i K_{z} V(r)] B_{r}(r)  = 
\int_{0}^{\infty} d \xi F(\xi) e^{-[\gamma - i K_{z} V(r)] \xi} 
\nonumber \\
\times \left[ \frac{d^{2} B_{r}}{d r^{2}} - K_{\perp}^{2} B_{r} + i \xi 
V'(r) K_{z} \frac{d B_{r}}{dr} \right],
\end{eqnarray}
where $K_{\perp}^{2}= K_{z}^{2} + M^{2}/r^{2}$. Note that the radial
dependence of the azimuthal component of $\vec{B}$ satisfies the same
equation in our model. Integrating Eq.~(24) over $d \xi$, we have
\begin{equation}
\frac{d^{2} B_{r}}{d r^{2}} - i K_{z} V'(r) \frac{\lambda_{T}}{\mu_{T}}
\frac{d B_{r}}{d r} - \left( K_{\perp}^{2} + \frac{\Gamma}{\mu_{T}} 
\right) B_{r} = 0
\end{equation}
where 
\begin{equation}
\mu_{T} = \frac{4 \pi}{3} \; \Gamma \int 
\frac{G(\omega, \vec{k})}{\omega^{2} + \Gamma^{2}} 
k^{4} d \omega dk,
\end{equation}
\begin{equation}
\lambda_{T} = \frac{4 \pi}{3} \int \frac{(\omega^{2} - \Gamma^{2})
G(\omega, \vec{k})}{(\omega^{2} + \Gamma^{2})^{2}} k^{4} d \omega
dk,
\end{equation}
and $\Gamma = \Gamma(r)= \gamma - iK_{z} V(r)$. The coefficient 
$\mu_{T}$ represents a nonlocal magnetic viscosity in a turbulent 
shear flow, and the coefficient $\lambda_{T}$ describes a qualitatively 
new turbulent kinetic process that can be responsible for the 
generation of the mean field. Note that, in our nonlocal model, 
turbulent kinetic coefficients depend on the rate of a mean process 
that is the principle difference to any local theory like a two-scale 
approximation. The kinetic coefficients are complex in our model
since $\Gamma$ is complex. It is convenient to represent $B_{r}(r)$ 
as
\begin{equation}
B_{r}(r) = f(r) \exp \left( \frac{i}{2} K_{z} \int V'(r') 
\frac{\lambda_{T}}{\mu_{T}} d r' \right).
\end{equation}
Then, the equation for $f(r)$ reads
\begin{eqnarray}
\frac{d^{2} f}{ d r^{2}} - \left[K_{\perp}^{2} + \frac{\Gamma}{\mu_{T}}
- \frac{1}{4} K_{z}^{2} V'^{2}(r) \frac{\lambda_{T}^{2}}{\mu_{T}^{2}}
\right.
\nonumber \\
\left. - \frac{i}{2} K_{z} \frac{d}{dr} \left( V'(r) 
\frac{\lambda_{T}}{\mu_{T}} \right) \right] f = 0.
\end{eqnarray}

The coefficients $\mu_{T}$ and $\lambda_{T}$ depend on $\Gamma$ and, 
hence, on $r$, so these dependences are determined by the spectral 
function. As mentioned, the origin of turbulent motions, as well 
as their spectrum, are rather uncertain in jets. Turbulent motions can 
be caused, for example, by acoustic waves generated due to the 
Kelvin-Helmholtz instability (see, e.g., Payne \& Cohn 1985). In this 
paper, we assume as a certainty that turbulence is acoustic and choose 
the simplest possible dependence of $G(\omega, \vec{k})$ on $\omega$,
\begin{equation}
G(\omega, \vec{k}) = \frac{G(\vec{k})}{\omega^{2} + 
\tau^{-2}}, 
\end{equation}  
where $\tau$ is the characteristic correlation timescale of
turbulence. Dependence (30) corresponds to a velocity correlation 
tensor exponentially decreasing with time, 
\begin{equation}
\langle v_{i}(\vec{r}, t) v_{j}(\vec{r}, t+ \Delta t) \rangle
\propto e^{- \Delta t/\tau}
\end{equation} 
(see, e.g., R\"{u}diger 1989). Then, the kinetic coefficients are
\begin{equation}
\mu_{T} = \frac{\tau v_{T}^{2}}{1 + \Gamma \tau}, \;\;\;\;
\lambda_{T} =  - \frac{\tau^{2} v_{T}^{2}}{(1 + \Gamma \tau )^{2}},
\end{equation}
where $v_{T}$ is the characteristic turbulent velocity,
\begin{equation}
v_{T}^{2} = \frac{8}{9} \pi^{2} \tau \int G(\vec{k}) k^{4} dk.
\end{equation}

Substituting expressions (32) into Eq.~(29) and taking into 
account that $\Gamma' = -i K_{z} V'$,  we obtain 
\begin{equation}
\frac{d^{2} f}{d r^{2}} + q^{2}(r) f =0,
\end{equation}
where
\begin{equation}
q^{2}(r)= - \frac{\Gamma}{\mu_{T}}
- K_{\perp}^{2} + \frac{3}{4} \frac{K_{z}^{2} \tau^{2} 
V'^{2}}{(1 + \Gamma \tau)^{2}}  
- \frac{i}{2} \frac{K_{z} \tau V''}{1 + \Gamma \tau}.
\end{equation}
To solve Eq.~(34) one needs the corresponding boundary conditions. 
For the sake of simplicity, we assume that plasma inside and outside 
of the shear layer is highly conductive and that the mean-field does not 
penetrate into the surrounding medium. Then, function $f(r)$ 
should be vanishing at $r=R$ and $r=R-\Delta r$. The main qualitative 
conclusions are the same for other possible boundary conditions. 

To estimate the eigenvalues of Eq.~(34), we can use an integral
method similar to the one proposed by Chandrasekhar (1960). Since $f$ is 
complex, we multiply Eq.~(34) by the complex conjugate function 
$f^{*}$ and integrate over the whole region where the magnetic field 
is generated. Then, we have
\begin{equation}
\int_{R - \Delta r}^{R} \left| \frac{df}{dr} \right|^{2} dr -
\int_{R - \Delta r}^{R} q^{2}(r) \left| f \right|^{2} dr =0 
\end{equation} 
for the chosen boundary conditions. Both $|df/dr|^{2}$ and $|f|^{2}$ 
are positive quantities over the integration domain, but $q^{2}(r)$ is 
complex. Splitting Eq.~(36) into the real and imaginary parts, we 
obtain
\begin{eqnarray}
\int_{R-\Delta r}^{R} \left| \frac{df}{dr} \right|^{2} dr =
\int_{R-\Delta r}^{R} \mathrm{Re} \; q^{2}(r) 
\left| f \right|^{2} dr, \\
\int_{R-\Delta r}^{R} \mathrm{Im} \; q^{2}(r) \left| f \right|^{2} dr 
=0.
\end{eqnarray}
By applying the mean value theorem, Eqs.~(37) and (38) can be transformed
into 
\begin{eqnarray}
\lefteqn{ \int^{R}_{R-\Delta r} \mathrm{Re} \; q^{2}(r) dr = 
\frac{\Delta r}{(\Delta R)^{2}},} \\
\lefteqn{\left| f(r_{2}) \right|^{2} \int_{R - \Delta r}^{R} 
\mathrm{Im} \; q^{2}(r) dr =0,} 
\end{eqnarray}
where 
\begin{equation}
\frac{\Delta r}{(\Delta R)^{2}} = \int_{R - \Delta r}^{R} \left| 
\frac{df}{dr} 
\right|^{2} dr \Big/ \left| f(r_{1}) \right|^{-2}, 
\end{equation}
$\Delta R \sim \Delta r$ is the characteristic radial lengthscale of 
the magnetic field, and $r_{1}$ and $r_{2}$ are some mean points within 
the shear layer. Equations (39) and (40) can be combined into
\begin{equation}
\int^{R}_{R-\Delta r} q^{2}(r) dr = 
\frac{\Delta r}{(\Delta R)^{2}}.
\end{equation}
Substituting expression (35), we obtain the dispersion equation
for dynamo modes
\begin{eqnarray}
\int^{R}_{R-\Delta r} \! \left[ \Gamma \tau (1 + \Gamma \tau) \!-\! 
\frac{3 K_{z}^{2} \ell^{2} \tau^{2} V'^{2}}{4 (1 + \Gamma \tau)^{2}} 
+ \frac{i K_{z} \ell^{2} \tau V''}{2 (1 + \Gamma \tau)} \right] dr 
\nonumber \\
= -\ell^{2} Q^{2} \Delta r,
\end{eqnarray}
where $V''=d^{2} V/dr^{2}$, and
\begin{equation}
Q^{2} \!= \! \frac{1}{(\Delta R)^{2}} + \frac{1}{\Delta r} 
\int_{R-\Delta r}^{R} K_{\perp}^{2} dr \approx \frac{1}{(\Delta R)^{2}}
+ K_{z}^{2} + \frac{M^{2}}{R^{2}}, 
\end{equation}
where $Q$ is the characteristic wavevector of dynamo waves,
and $\ell= \tau v_{T}$ is the lengthscale of turbulence.

To estimate the eigenvalues of Eq.~(43), we initially consider 
the simplest model assuming that shear is approximately linear within 
the shear layer, $R \geq r \geq R-\Delta r$. 
In this case, the mean velocity can be represented as
\begin{equation}
V(r) = V_{0}(R-r)/\Delta r,
\end{equation} 
where $V_{0}$ is the velocity in the jet core, $r< R-\Delta r$. Then, 
Eq.~(43) yields
\begin{equation}
i\int^{\tau \Gamma_{e}}_{\tau \Gamma_{i}} \left[ x 
(1 + x) - \frac{3 K_{z}^{2} \ell^{2} \tau^{2} V'^{2}}{4 
(1 + x)^{2}} \right] dx 
= K_{z} V_{0} \tau \ell^{2} Q^{2},
\end{equation}
where $\Gamma_{e} \equiv \Gamma(R) = \gamma$ and $\Gamma_{i} \equiv
\Gamma(R-\Delta R)=\gamma - i K_{z}V_{0}$ are the values of $\Gamma$
at the outer and inner boundaries of the shear layer. Integrating 
Eq.~(46), we obtain
\begin{eqnarray}
(1 + \gamma \tau)(\gamma - i K_{z} V_{0}) + \frac{i}{2} K_{z} V_{0} 
+ \frac{\ell^{2}}{\tau} Q^{2} -
\nonumber \\
\frac{\tau}{3} K_{z}^{2} V_{0}^{2} \left[ 1 + 
\frac{(3 \ell/2 \Delta R)^{2}}{(1 + \gamma \tau)(1 + \gamma \tau -
i \tau K_{z} V_{0})} \right] = 0. 
\end{eqnarray}
The term proportional to $(3 \ell/2 \Delta R)^{2}$ is small in the
mean-field theory and can be neglected compared to 1 in Eq.~(47). Then, 
the dispersion equation simplifies
\begin{equation}
(1 + \gamma \tau)(\gamma - i K_{z} V_{0}) + \frac{i}{2} K_{z} V_{0} 
- \frac{\tau}{3} K_{z}^{2} V_{0}^{2} + \frac{\ell^{2}}{\tau} Q^{2} 
= 0. 
\end{equation}
Splitting $\gamma$ into real and imaginary parts, $\gamma = i \gamma_{I}
+ \gamma_{R}$, we obtain two equations for $\gamma_{I}$ and $\gamma_{R}$ 
from Eq.~(48)
\begin{equation}
(1+ 2 \gamma_{R} \tau) \left( \gamma_{I} - \frac{1}{2} K_{z} V_{0} 
\right) =0,
\end{equation}
\begin{equation}
\gamma_{R} (1 + \gamma_{R} \tau) - \gamma_{I} \tau (\gamma_{I} -
K_{z} V_{0} ) - \frac{\tau}{3} K_{z}^{2} V_{0}^{2} + 
\frac{\ell^{2}}{\tau} Q^{2} = 0.
\end{equation}
Equation (49) yields
\begin{equation}
\gamma_{I} = \frac{1}{2} K_{z} V_{0}.
\end{equation}
Then, we have from Eq.~(50)
\begin{equation}
(\tau \gamma_{R})^{2} + \tau \gamma_{R} - 
\left( \frac{\tau^{2}}{12}
K_{z}^{2} V_{0}^{2} - \ell^{2} Q^{2} \right) = 0.
\end{equation}
The roots of this equation are
\begin{equation}
\tau \gamma_{R \; 1,2} = - \frac{1}{2} \pm \sqrt{ \frac{1}{4} +
\left( \frac{\tau^{2}}{12}
K_{z}^{2} V_{0}^{2} - \ell^{2} Q^{2} \right)}.
\end{equation}
One of the roots is positive, so the corresponding dynamo
mode is growing if
\begin{equation}
\frac{\tau^{2}}{12} K_{z}^{2} V_{0}^{2} > \ell^{2} Q^{2}.
\end{equation}
Estimating $Q^{2} \approx 1/ (\Delta R)^{2} + K_{z}^{2}$ and assuming 
that the azimuthal wavelength is much larger than $\Delta R$, we can 
represent Eq.~(54) as
\begin{equation}
K_{z} \Delta R \; \frac{V_{0}}{v_{T}} > 2 \sqrt{3} \left(1 - 
\frac{12 v_{T}^{2}}{V_{0}^{2}} \right)^{-1/2}.
\end{equation}
This condition cannot be fulfilled if $V_{0} < 2 \sqrt{3} v_{T}$,
and the dynamo does not operate in such jets. However, it is plausible 
that the mean velocity is much larger than the turbulent velocity in jets,
$V_{0} \gg 2 \sqrt{3} v_{T}$, and condition (55) can be satisfied for 
a wide range of dynamo modes with
\begin{equation}
\lambda_{z} < \frac{\pi V_{0}}{\sqrt{3} v_{T}} \; \Delta R,
\end{equation}
where $\lambda_{z}=2 \pi/ K_{z}$ is the wavelength in the $z$-direction.
Therefore, the maximum longitudinal wavelength of the dynamo-generated
magnetic structure in jets is of the order of $\Delta R (V_{0}/v_{T})$. 

The generation time, $t_{*} = 1/ \gamma_{R}$, is given by
\begin{equation}
t_{*} = 2 \tau \left[ -1 + \sqrt{ 1 + \frac{K_{z}^{2} \tau^{2}}{3}
(V_{0}^{2} - 12 v_{T}^{2}) - \frac{4 \ell^{2}}{(\Delta R)^{2}} }
\right]^{-1}.
\end{equation}
For dynamo waves with a relatively short wavelength, $\lambda_{z}
< 2 \pi \ell (V_{0}/ v_{T})$ (or $V_{0} \gg 1 / \tau K_{z}$), the 
growth time is
\begin{equation}
t_{*} \sim \frac{2 \sqrt{3}}{K_{z} V_{0}}.
\end{equation}
In the limit of a large wavelength, $\lambda_{z} > 2 \pi \ell
(V_{0}/ v_{T})$ (but still satisfying condition (55)), Eq.~(57) yields
\begin{equation}
t_{*} \sim \frac{12 \tau}{K_{z}^{2} \ell^{2}} \; 
\frac{v_{T}^{2}}{V_{0}^{2}}. 
\end{equation}
The dynamo modes with the maximum possible wavelength ($\sim \Delta R
(V_{0}/v_{T})$) grow on the slowest timescale $t_{*} \sim \tau
(\Delta R/\ell)^{2}$.

It is seen from this consideration that generation of the mean-field 
is determined by a velocity difference between the boundaries of a shear
layer rather than by the details of the velocity profile. Therefore, our 
results can be generalized for any velocity profile in a relatively 
simple way. We can split Eq.~(43) into the real and imaginary parts and 
then obtain equations analogous to Eqs.~(51) and (52) by applying the 
mean value theorem. The only difference to Eqs.~(51) and (52) is that 
the equations for a more complicated velocity profile will contain
the value of a flow velocity on some mean point $V(r_{*})$ ($R > r_{*}
> R -\Delta r$) instead of $V_{0}$. Correspondingly, the growth rate 
and the generation condition will be given by Eqs.~(53) and (54), 
respectively, with the replacement $V_{0} \rightarrow V(r_{*})$.
Therefore, a particular shape of the velocity profile appears to be 
unimportant in our model, and large-scale magnetic fields can be
generated in any jets with a strong shear.

\section{Discussion}

We have considered the turbulent dynamo action in jets. The main 
result is that even the simplest turbulent motions showing the mirror 
symmetry become suitable for the generation of a large-scale magnetic 
field in the presence of shear. An amplification of 
the mean field takes place due to non-local terms that appear in the 
mean electromotive force and are caused by shear stresses. The 
considered mechanism of generation is qualitatively different from 
the conventional turbulent alpha-dynamo that, apart from the lack of 
the mirror symmetry of turbulence, also requires large-scale 
stratification. Due to its simplicity, the proposed mechanism is well 
adopted to the physical conditions in jets because the presence of 
both shear and turbulence seems to be plausible in a jet flow. 

Unfortunately, neither available observational data nor theoretical 
modelling provide reliable information concerning the velocity 
profile in jets. However, the generation of a large-scale magnetic 
field can take place for any velocity profile, which is an attractive
feature of our model. The only necessary condition of the considered 
dynamo is the presence of a sufficiently strong shear satisfying 
condition (54). This condition can be fulfilled in many jets or, 
at least, in a fraction of their volume. This dynamo 
mechanism generates the field in jets on a very short timescale that 
can be comparable to the turnover time of turbulence $\tau$ and that
is typically much shorter than the lifetime of jets. Therefore, the 
generation can most likely reach a saturation level when the dynamo 
works in the nonlinear regime.

We considered generation of the radial component of $\vec{B}$ since
this component is the most important one in the shear-driven dynamo.
Generally, two other components can be stronger but their evolution
is determined entirely by the behaviour of $B_{r}$. We can estimate 
$B_{z}$ from the $z$-component of the mean induction equation, equating 
the dissipative and stretching terms. Then,
\begin{equation}
B_{z} \sim \frac{V' (\Delta R)^{2}}{\mu_{T}} B_{r} \sim
K_{z} \Delta R \left(\frac{V}{v_{T}} \right)^{2} \; B_{r},
\end{equation}
and the longitudinal magnetic field is always stronger than the 
radial one. The azimuthal magnetic field can be estimated from 
the divergence condition and can vary within a wide range depending 
on the longitudinal wavelength and $M$. For not very large $M \neq 0$, 
$B_{\varphi}$ is typically stronger than the radial field. Our 
linear analysis does not allow proper estimation of the saturation
magnetic field but, most likely, the strongest field component can 
reach equipartition with the kinetic energy of jet particles.

In our model, a dynamo generates a large-scale field in the layer
of the thickness $\Delta r$ near the jet surface. The shear-driven 
dynamo is much less efficient in the core region, $r < R- \Delta r$.
However, if the jet is turbulent, the small-scale turbulent dynamo 
can amplify small-scale magnetic fields in the core region even if 
the shear is negligible. The magnetic Reynolds number is large in jets, 
and turbulent motions caused, for example, by instabilities stretch 
and distort the field lines increasing thereby the energy of 
generated small-scale magnetic fields rapidly (see, e.g., Schekochihin et al. 
2001). This random field reaches equipartition with the energy of 
turbulent motions on a very short timescale $\sim \tau$. 
Since turbulent motions are less energetic than the mean flow, we can 
expect that random fields in the central region are typically weaker 
than the large-scale magnetic field in a magnetic sheath surrounding 
the jet. The characteristic lengthscale of turbulent 
magnetic fields in the central region can be shorter than $\ell$.

{}
\end{document}